# An agent-based modeling approach for real-world economic systems: Example and calibration with a Social Accounting Matrix of Spain

Martin Jaraiz

Department of Electronica, University of Valladolid, 47011 Valladolid, Spain.

email: mjaraiz@uva.es

The global economy is one of today's major challenges, with increasing relevance in recent decades. A frequent observation by policy makers is the lack of tools that help at least to understand, if not predict, economic crises. Currently, macroeconomic modeling is dominated by Dynamic Stochastic General Equilibrium (DSGE) models. The limitations of DSGE in coping with the complexity of today's global economy are often recognized and are the subject of intense research to find possible solutions. As an alternative or complement to DSGE, the last two decades have seen the rise of agent-based models (ABM). An attractive feature of ABM is that it can model very complex systems because it is a bottom-up approach that can describe the specific behavior of heterogeneous agents. The main obstacle, however, is the large number of parameters that need to be known or calibrated. To enable the use of ABM with data from the real-world economy, this paper describes an agent-based macroeconomic modeling approach that can read a Social Accounting Matrix (SAM) and deploy from scratch an economic system (labor, activity sectors operating as firms, a central bank, the government, external sectors...) whose structure and activity produce a SAM with values very close to those of the actual SAM snapshot. This approach paves the way for unleashing the expected high performance of ABM models to deal with the complexities of current global macroeconomics, including other layers of interest like ecology, epidemiology, or social networks among others.

## 1. Introduction

A recurring complaint of policy makers is the lack of useful macroeconomic models and tools: ”When the crisis came, the serious limitations of existing economic and financial models immediately became apparent. (...) Macro models failed to predict the crisis and seemed incapable of explaining what was happening to the economy in a convincing manner. As a policymaker during the crisis, I found the available models of limited help. In fact, I would go further: in the face of the crisis, we felt abandoned by conventional tools.” [1]

At the time when John M. Keynes [2] outlined his ideas for a systematic modeling of macroeconomics, models were analytical in their simplicity, describing only the main guidelines of the economy. An economic system can be modeled as a dynamic structure ("hardware": households, firms, banks, government...) driven by behavioral rules ("software": supply-demand, pricing, fiscal and monetary policies, trusts...) [3]. With the advent of computers, the search for more detailed and accurate models led to a sustained effort to improve mainly the behavioral rules, neglecting the key role of a more detailed representation of the structure of the economic system. This bias to focus on rules may have been fostered by the fact that, in the analytical modeling approach, increased disaggregation soon leads to a prohibitive computational burden.

To some extent, a complex macroeconomic system is constantly evolving, like a living organism, and the field of economics can be compared to that of medicine. In the 19th century, both physicians and economists were guided primarily by their own experience and skill. Today, however, any physician can rely on the help of state-of-the-art analytical instruments and data to help identify which organ is causing the disease and take quantitative measures to return the organism to a healthy state. In contrast, economists have experienced little progress in the tools available to address macroeconomics as a quantitative science. While microeconomics can still be tracked and analyzed in detail thanks to computers, macroeconomic systems are now entangled in networks of interaction with a spatial scope and temporal pace that cannot be accurately modeled unless the system as a whole is modeled. Just as mechanical engineering once moved from the steam engine to the combustion engine to the electric motor, the time may have come for economics to move to alternative modeling techniques that take full advantage of the computational power available today. However, despite this computational power, dynamic stochastic general equilibrium (DSGE) models cannot account for today's global economy because the complexity of the network of interactions makes the economic model intractable for equation-based modeling schemes.

Another approach, agent-based modeling (ABM) [4], can easily and effectively handle highly disaggregated structures. It has been proposed as a tool to be explored [1]: “The atomistic, optimising agents underlying existing models do not capture behaviour during a crisis period. We need to deal better with heterogeneity across agents and the interaction among those heterogeneous agents. (...) Agent-based modeling dispenses with the optimisation assumption and allows for more complex interactions between agents. Such approaches are worthy of our attention.”

In fact, several agent-based macroeconomic modeling environments have already been developed; see [5] for a review. However, most of them have been used to conduct research in different specific areas of economic policy, and not as a general simulator of the real economy of a region or set of regions such as, for example, the European Union. Strikingly, none of the four main valid models for the euro area recently listed by [6] is agent-based. This work presents an agent-based approach that could, eventually, tackle the task of simulating the global real-world macroeconomy with highly disaggregated heterogeneous agents.

A driving force for developing this approach is the assumption of the critical role of a detailed structure of the macroeconomic system. Using an analogy with road traffic, it is to be expected that the behavioral rules of drivers will be similar in, for example, Madrid or Paris. However, to assess or design routes to alleviate traffic congestion in those cities, it is essential to have a sufficiently disaggregated road map, with not only the outline of the main industrial, commercial, and residential areas, but also their internal roads, squares and junctions. To implement a realistic traffic simulator, knowledge of this more detailed structure, supplemented by a small set of simple behavioral rules, may be more useful than a simplistic black box structure with complex behavioral rules. Similarly, one would expect that the main difference between Madrid and Paris, considered as macroeconomic systems, or between Spain and France, would arise from differences in their economic structures, rather than from the behaviors of their agents: given the economic fabric of a country, even the best business managers do not have much room to improve the economy in the short term, until the economic structure is improved, because they are constrained by their neighborhood of interaction.

And, returning to the example of medicine, today the goal and success of doctors and hospitals is not to predict illness in healthy people, but to cure those who are already sick or injured. In macroeconomics, a complaint of policy makers is that current tools "seemed unable to explain what was happening to the economy in a convincing way" [1] during the crisis. Understanding what is happening is the first step in devising a possible solution or a warning of an impending crisis. The degree of structural and functional detail now available to medicine may be one of the major contributions to its progress and success in the last century.

Most developed countries already have access to a large amount of data on their own economic structure and can easily collect more data if needed. The stumbling block, however, is the lack of suitable software that can take this amount of data and simulate the flow of wealth as, for example, some Internet mapping applications do with global road traffic, not to the minute as in traffic, but on a time scale of a month or a quarter. One of the objectives of the line of work presented here is to explore whether this is possible with a simulation scheme based on self-deploying agents [3], adapted to replicate the economy of a region, or set of regions, using for example data from a Social Accounting Matrix (SAM) for calibration. It should be noted that this work does not propose yet another ABM model, but rather a methodology to implement any ABM model so that the model deploys an economic system that meets the desired calibration data. And it does so by taking advantage of the self-adaptive capabilities of economic systems driven by supply-demand and competition (survival of the fittest).

Specifically, the methodology presented here is an attempt to develop an agent-based macroeconomic modeling approach that can read a Social Accounting Matrix (SAM), Fig. 1, and deploy from scratch an economic system (labor, activity sectors operating as firms, a central bank, the government, external sectors...) whose structure and activity yield a SAM with values in close agreement with those of the actual SAM snapshot. Subsequent evolution depends on the behavioral rules defined by the modeler, which are applied from the beginning, once the system is allowed to evolve without constraints after an initial deployment and calibration phase. Thanks to the plasticity of economic systems, which can modify their structure through the demand-supply mechanism to respond to endogenous or exogenous stimuli, the behavioral rules can also be time-dependent.

The purpose of this short paper is to give some guidelines of the modeling approach to allow its application by other groups; a more detailed description will be published elsewhere. For the simple behavioral rules included in the example, two thousand (active) individuals are sufficient to obtain statistically acceptable results and it takes about two minutes on standard CPUs. Since agents can have a limited number of interactions per month, the CPU time increases linearly with the number of simulated individuals.

## 2. Description of the model

The framework is a follow-up of the one described in Ref. [3]. As a simple example, we have chosen a SAM of Spain [7] with only six activity sectors and implemented a set of basic behavioral rules for the agents. As indicated above, the point of interest is not the agent-based model used for this example, but the capability of self-deployment of an economic structure, built with that model, that yields a SAM with values close to the desired SAM data.

The implemented model simulates the activity of a population of households (active individuals) in a geographic region, initially with a government, a central bank (CB) and an external sector. As the simulation progresses, some households will open firms, of types based on local demand for goods, and some commercial banks, and a financial market will start as some firms reach a sufficiently large net worth (or meet any other requirement). Unprofitable companies close and their owners look for work in the neighborhood. Prices evolve from interaction with neighbors following simple supply and demand rules: if the buyer's price is greater than or equal to the seller's price, the transaction occurs, and the buyer's price decreases and the seller's price increases by a small factor. Otherwise, there is no transaction and prices change in the opposite direction.

Households initially have some monthly cash wages. They go out once every time step (month) on a random day to buy goods from the neighborhood, initially in the relative proportion given by the SAM (households consumption) but modulated with a logit probability as a function of prices [4]. The consumption budget at step t, $C_{h,t}$, depends on the average income ($I_{h,t}$) and wealth ($W_{h,t}$) of each household, given by the following "buffer stock" rule [4]:

$$C_{h,t} = I_{h,t} + \kappa \cdot (W_{h,t} - \varphi \cdot I_{h,t}) \qquad [1]$$

where $\kappa$ is a sensitivity parameter and $\varphi$ a buffer size. Households divide their surplus (if any) into bank deposits and risky assets (equity shares of individual firms) in the stock market.

Initially there are no firms, but households without a firm have a certain probability of opening a new one of one of the most frequently ordered types in the neighborhood. A firm produces only one type of homogeneous good whose unit is the quantity that could be purchased with one monetary unit (mu) in the SAM year. Therefore, initially all prices are 1.0. Unlike households, firms can borrow from banks (CB until the founding of the first commercial bank) and have a random but fixed day of the month for their productive activity. They can sell out of their stock on any day to households or to other firms for intermediate consumption (IC). Each firm attempts to produce enough to replenish a level of inventories that is estimated based on recent demand. If the firm has liquidity needs to finance production (IC and taxes according to its SAM column ratios, plus labor and capital according to a Cobb-Douglas or Leontief model) it can apply for a bank loan and the subsequent production volume is conditioned to the outcome of the application. In addition, if a company meets the requirements to enter the stock market, it can also issue new shares. After taxes are paid, a fraction of the profits is distributed as dividends to the owner or shareholders, and the remainder is deposited in the payment account. Consistent with what was said above about the primary role of a policymaker-oriented macroeconomic model as a tool for a quantitative understanding of the short-run functioning and response of the system, we have not introduced capital goods additions or the effects of new technologies or skills. Those issues are postponed to the more ambitious goal of predicting medium- and long-term behavior, if this modest short-term prediction turns out to be sufficiently reliable and accurate to be useful.

In turn, capital goods are produced as consumption goods: a new door can be used as gross fixed capital formation (GFCF) for a new building or as a consumption good to replace another door in an old building. We use the coefficients in the GFCF column of the SAM to divide the GFCF row value of, for example, the Households column into its goods and services components. In this way, the contribution of GFCF output to the economy is fully accounted for.

The financial market takes place in a clearing house which collects at the end of the month all the sell and buy orders (sorted from high to low price) and allocates them starting with the high price orders. Prices are readjusted following the supply and demand rule mentioned above.

Banks can be founded by families that meet the criteria defined by the Central Bank, such as a minimum initial net worth and a maximum number of banks in total. We have followed as a guide the model applied by Dawid et al. [4], in which the bank's ability to extend credit is constrained by a capital adequacy requirement (CAR) and a reserve requirement ratio (RRR), but with a simple first-come, first-served response to loan applicants. The interest rate offered to a firm is an increasing function of credit risk, based on the firm's probability of default on the loan, which is estimated from its debt-to-equity ratio.

The government, according to the data provided by the SAM, collects taxes and redistributes them in the form of subsidies (unemployment) to households and public spending. In our simple model, the government uses the Central Bank to deposit or withdraw its surplus or deficit, respectively.

The external sector, like the government, is implemented as a simple input-output or a consumer (its SAM column) - producer (its SAM row of IC for firms) account, but with unlimited cash instead of using the Central Bank.

## 3. Self-deployment approach and results

The deployment of the economic system is based on a simple survival-of-the-fittest approach. Initially there are no firms, only the active population, the government, the CB, and the external sector. To build up the production structure (number, fixed capital size, number of employees, and activity level of firms), it is initially established that the final consuming agents (households, government, and external sectors) repeatedly try to buy from neighbors their SAM monthly value. During an initial deployment stage (up to month 360 in this example), individuals open firms corresponding to activity sectors 1 to 6 of the SAM, and close unprofitable ones, until the production levels match the experimental values (output, unemployment, final and intermediate consumption). Producers also try to buy from the firms in their neighborhood the intermediate consumption goods (IC) they need. The deployment stage ends when the economic system reaches a steady state with the desired level of activity.

This configuration snapshot (deployed economy) can be saved as an initial state to run different what-if tests. From this initial state on, the system is allowed to evolve according to the demand, supply and behavioral rules established by the model, without imposing the SAM's consumption quantities. For example, each household consumption budget follows eq. [1].

As Fig. 2 (a) shows, initially the entire labor force is unemployed, but new firms start to open and hire employees according to the needs of each sector, Fig. 2 (b), to satisfy the demand of the final consumers, Figs. 2 (c), (d), (e), plus the IC demanded by other firms, Fig. 2 (f). From step 240 to 360 the economy shows a steady state, Figs. 2 (a) to (g), including the firms' inventories of IC and goods for sale. However, the wealth of households increases, Fig. 2 (h). This is because the household IS value of the SAM is positive. For this reason, their consumption budget, eq. [1], has been adjusted up to step 360 by a factor, to keep consumption levels equal to the SAM values. At the end of the deployment phase, this factor is locked to its last value and household consumption starts to increase, Fig. 2(c), while the government and the external sector maintain their own consumption levels, Figs. 2 (d) and (e).

The deployed economy can then be analyzed and compared with, for example, SAM data. Fig. 3 shows the intermediate and final consumption values from the simulation (step 360 of Fig. 2), as % of the actual SAM data. Small values in the SAM data yield large statistical errors in the simulation due to the discretization of employees and firms.

Figure 3 is a comparison of the distribution of household wealth from the simulation, with the simple model given above, and the distribution of wealth in Spain in 2021.

Finally, Fig. 5, corresponding to another SAM of 42 activity sectors, shows that the program can deploy economic systems without the user having to adjust any internal parameters for convergence.

## 1. Conclusion

The described methodology enables the use of agent-based models to real-world economies, as shown with a basic model calibrated to replicate two SAMs of Spain in 2008. Although these are just proof-of-concept results, it opens the way to the use of ABM as a tool for policy makers dealing with complex data from real macroeconomic systems. In addition, agent-based modeling can easily integrate different overlapping layers of interest like economics, ecology, epidemiology or social networks among others.

Competing interests: Author declares no competing interests.

| | A | B | C | D | E | F | G | H | I |
|---|---|---|---|---|---|---|---|---|---|
| 1 | SAM_table { MCAESP08 | | | | | | | | |
| 2 | SPAIN | Year: 2008 | | Population: 40000000 | | Active: 20000000 | | InitUnemp: 12 | |
| 3 | | P01_AgroPesc | P02_EnerPetro | P03_Indust | P04_Construc | P05_ServVenta | N06_ServNoVenta | F07_GFCF | X08_SectExt |
| 4 | P01_AgroPesc | 1701 | 1 | 24972 | 24 | 2877 | 302 | 811 | 7834 |
| 5 | P02_EnerPetro | 1119 | 41384 | 19205 | 2678 | 21061 | 6438 | 292 | 6740 |
| 6 | P03_Indust | 8616 | 5037 | 209653 | 64805 | 76816 | 21507 | 65355 | 144645 |
| 7 | P04_Construc | 190 | 793 | 1909 | 108372 | 25997 | 3654 | 176136 | 144 |
| 8 | P05_ServVenta | 3682 | 10762 | 94805 | 31726 | 244738 | 49376 | 54460 | 70689 |
| 9 | N06_ServNoVenta | 0 | 0 | 0 | 0 | 0 | 0 | 0 | 0 |
| 10 | F07_GFCF | 0 | 0 | 0 | 0 | 0 | 0 | 0 | 103916 |
| 11 | X08_SectExt | 8742 | 35259 | 238669 | 804 | 60405 | 1085 | 0 | 0 |
| 12 | L09_CompEmployees | 4259 | 4220 | 62958 | 50681 | 200109 | 88364 | 0 | 0 |
| 13 | K10_GrossOpSurplus | 19803 | 17740 | 48917 | 46113 | 321169 | 14029 | 0 | 0 |
| 14 | T11_SSoc | 624 | 1543 | 19007 | 15518 | 53763 | 26223 | 0 | 0 |
| 15 | T12_TaxProduction | -244 | 120 | -137 | 204 | 1118 | 53 | 0 | 0 |
| 16 | T13_TaxProducts | -471 | 307 | -1062 | 1555 | 12274 | 4100 | 21578 | -92 |
| 17 | T14_IRPF | 0 | 0 | 0 | 0 | 0 | 0 | 0 | 0 |
| 18 | G15_Government | 0 | 0 | 0 | 0 | 0 | 0 | 0 | 0 |
| 19 | H16_Households | 0 | 0 | 0 | 0 | 0 | 0 | 0 | 11088 |
| 20 | colSUM | 48021 | 117166 | 718896 | 322480 | 1020327 | 215131 | 318632 | 344964 |

| J | K | L | M | N | O | P | Q | R | S |
|---|---|---|---|---|---|---|---|---|---|
| Nproducers: 8 | | Naccounts: 16 | | Units: 1000000 | | euros | | | |
| L09_CompEmplo | K10_GrossOpSur | T11_SSoc | T12_TaxProduction | T13_TaxProducts | T14_IRPF | G15_Government | H16_Households | rowSUM | |
| 0 | 0 | 0 | 0 | 0 | 0 | 0 | 9499 | 48021 | P01_AgroPesc |
| 0 | 0 | 0 | 0 | 0 | 0 | 0 | 18249 | 117166 | P02_EnerPetro |
| 0 | 0 | 0 | 0 | 0 | 0 | 7620 | 114842 | 718896 | P03_Indust |
| 0 | 0 | 0 | 0 | 0 | 0 | 0 | 5285 | 322480 | P04_Construc |
| 0 | 0 | 0 | 0 | 0 | 0 | 21238 | 438851 | 1020327 | P05_ServVenta |
| 0 | 0 | 0 | 0 | 0 | 0 | 211972 | 3159 | 215131 | N06_ServNoVenta |
| 0 | 0 | 0 | 0 | 0 | 0 | 12014 | 202702 | 318632 | F07_GFCF |
| 0 | 0 | 0 | 0 | 0 | 0 | 0 | 0 | 344964 | X08_SectExt |
| 0 | 0 | 0 | 0 | 0 | 0 | 0 | 0 | 410591 | L09_CompEmployees |
| 0 | 0 | 0 | 0 | 0 | 0 | 0 | 0 | 467771 | K10_GrossOpSurplus |
| 0 | 0 | 0 | 0 | 0 | 0 | 0 | 20074 | 136752 | T11_SSoc |
| 0 | 0 | 0 | 0 | 0 | 0 | 0 | 0 | 1114 | T12_TaxProduction |
| 0 | 0 | 0 | 0 | 0 | 0 | 401 | 53758 | 92348 | T13_TaxProducts |
| 0 | 0 | 0 | 0 | 0 | 0 | 0 | 117483 | 117483 | T14_IRPF |
| 0 | 0 | 136752 | 1114 | 92348 | 117483 | 0 | 0 | 347697 | G15_Government |
| 410591 | 467771 | 0 | 0 | 0 | 0 | 94452 | 0 | 983902 | H16_Households |
| 410591 | 467771 | 136752 | 1114 | 92348 | 117483 | 347697 | 983902 | | |

**Figure 1**. A snapshot of a country's economy is, for example, this small Social Accounting Matrix (SAM) for Spain in 2008 [7]. Columns represent buyer (expenditures) and rows represent seller (receipts) accounts.

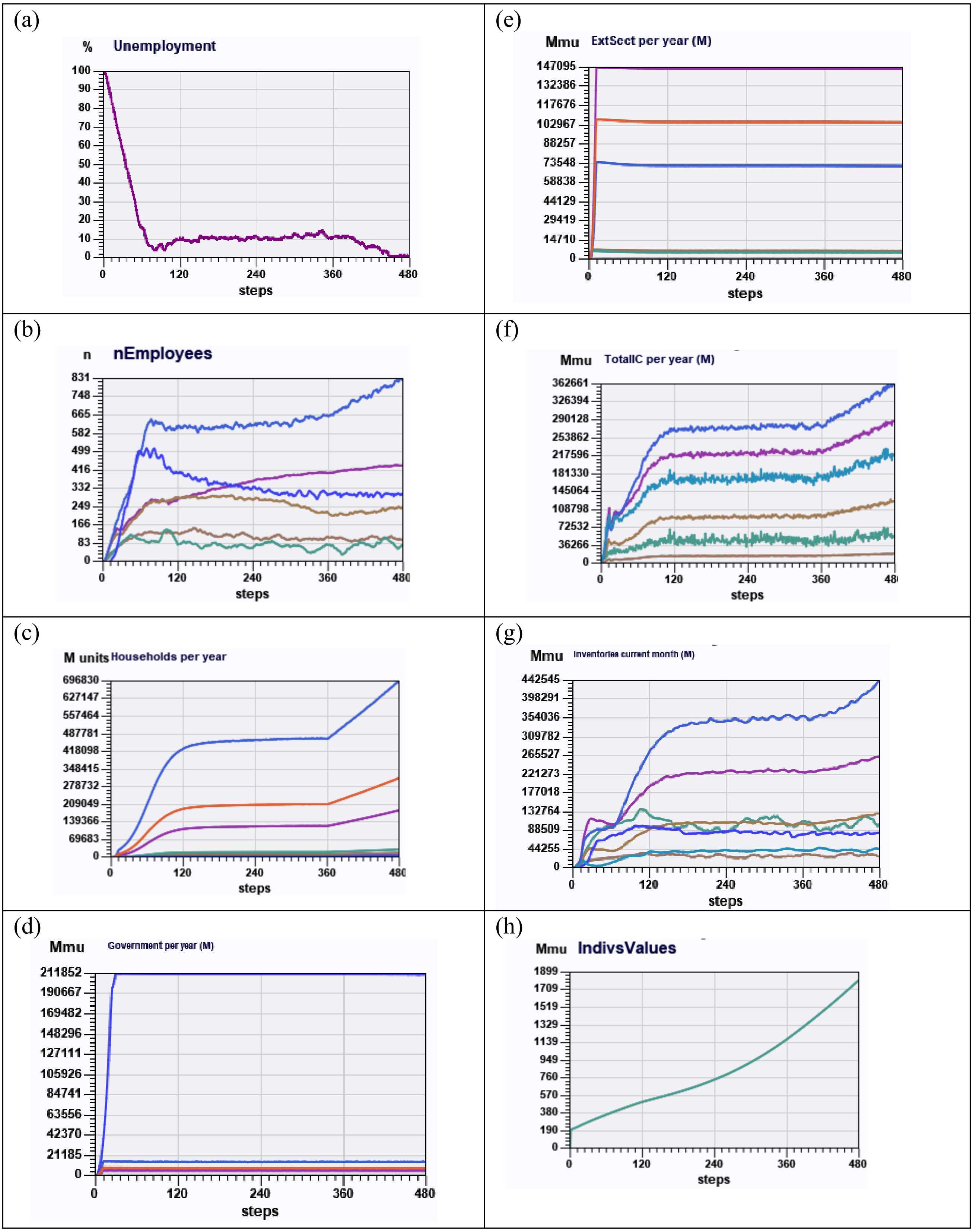

**Figure 2**. Temporal evolution of the deployment of the economy, until step 360, followed by free activity until step 480. (a) unemployment, (b) employees per sector, (c) households' consumption, (d) government, (e) external sector, (f) intermediate consumption, (g) inventories, and (h) households' wealth.

SAM calculated after absoluteStep 360: -----------------------------------------------------------

| | P01_AgroPes | P02_EnerPet | P03_Indust | P04_Constru | P05_ServVen | N06_ServNoV |
|---|---|---|---|---|---|---|
| P01_AgroPesc | 100 | 55 | 102 | 117 | 105 | 91 |
| P02_EnerPetro | 100 | 97 | 102 | 106 | 105 | 91 |
| P03_Indust | 100 | 97 | 102 | 106 | 105 | 91 |
| P04_Construc | 97 | 97 | 102 | 106 | 105 | 91 |
| P05_ServVenta | 100 | 97 | 102 | 106 | 105 | 91 |
| N06_ServNoVenta | | | | | | |
| F07_GFCF | | | | | | |
| X08_SectExt | 100 | 97 | 102 | 107 | 105 | 91 |
| L09_CompEmployees | 92 | 83 | 95 | 99 | 96 | 89 |
| K10_GrossOpSurplu | 100 | 92 | 104 | 109 | 105 | 98 |
| T11_SSoc | 100 | 92 | 104 | 109 | 105 | 98 |
| T12_TaxProduction | 100 | 92 | 104 | 109 | 105 | 98 |
| T13_TaxProducts | 100 | 92 | 104 | 109 | 105 | 98 |
| T14_IRPF | | | | | | |
| G15_Government | | | | | | |
| H16_Households | | | | | | |

| F07_GFCF | X08_SectExt | G15_Governm | H16_Househo | |
|---|---|---|---|---|
| 108 | 100 | | 105 | P01_AgroPesc |
| 102 | 100 | | 105 | P02_EnerPetro |
| 110 | 100 | 100 | 105 | P03_Indust |
| 111 | 100 | | 104 | P04_Construc |
| 110 | 100 | 100 | 105 | P05_ServVenta |
| | | 100 | 106 | N06_ServNoVenta |
| | 100 | 100 | 105 | F07_GFCF |
| | | | | X08_SectExt |
| | | | | L09_CompEmployees |
| | | | | K10_GrossOpSurplu |
| | | | 100 | T11_SSoc |
| | | | | T12_TaxProduction |
| | 100 | | 100 | T13_TaxProducts |
| | | | 100 | T14_IRPF |
| | | | | G15_Government |
| | | | | H16_Households |

**Figure 3**. Values of the simulated intermediate and final consumption and GFCF as % of the values of a 6 activity sectors SAM (missing values correspond to 0’s in the SAM) [7].

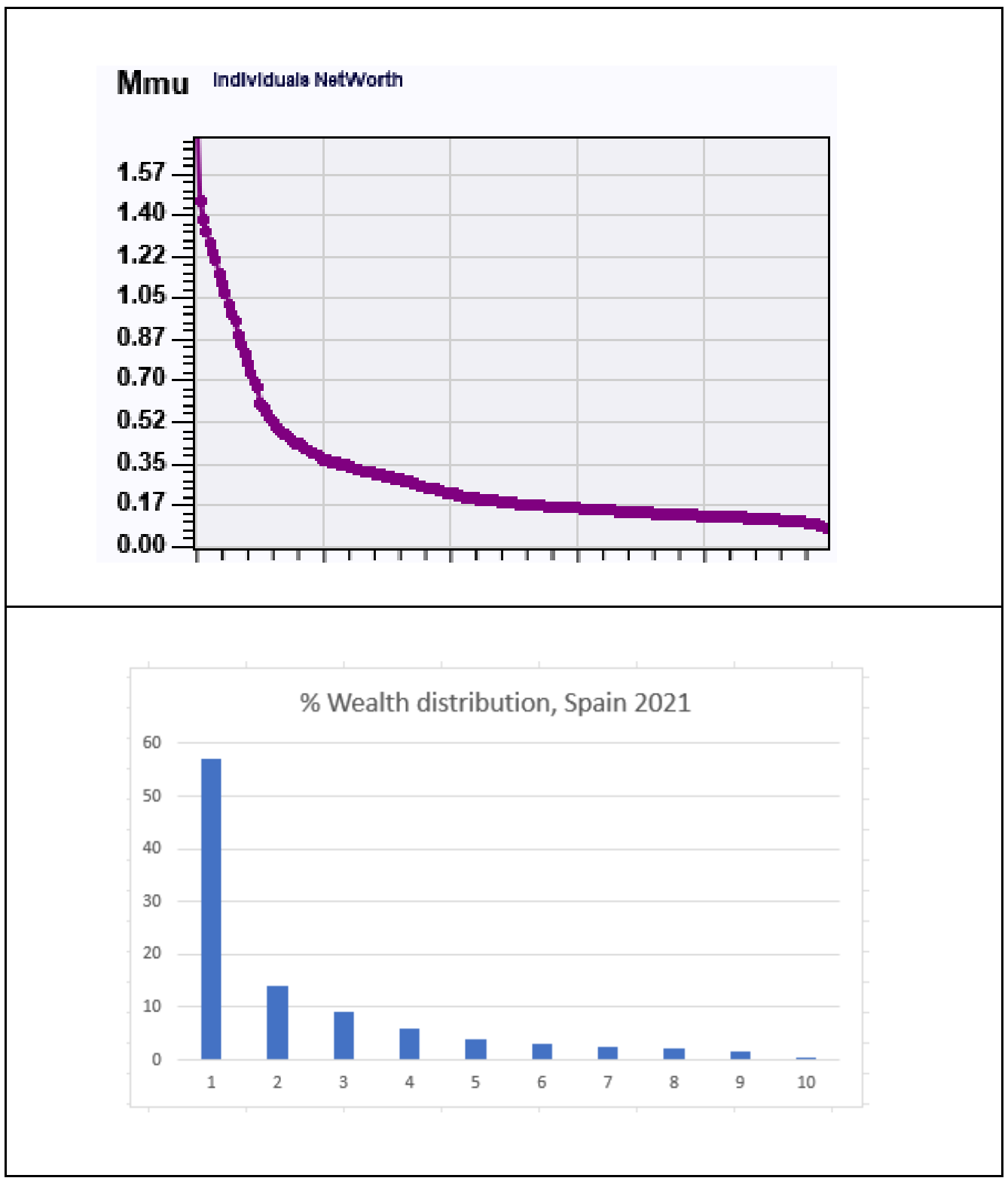

**Figure 4**. Comparison of the household wealth distribution from the simulation with the simple model used in the example (top) and the wealth distribution in Spain in 2021 (bottom).

| P14_Agua | P15_Aliment | P16_Textil | P17_FabricM | P18_Quimica | P19_MaterCo | P20_Metalur | P21_Metalic | P22_Maquina | P23_Vehicul |
|---|---|---|---|---|---|---|---|---|---|
| | 105 | 91 | 101 | 101 | | | | | 107 |
| | 103 | | | 110 | | | | | |
| | | | | 101 | 96 | 100 | | 205 | |
| | | | | 99 | | | | | |
| | 102 | 103 | 106 | 63 | 96 | 100 | 111 | 82 | 110 |
| 102 | 103 | 94 | 101 | 101 | 96 | 100 | 116 | 90 | 100 |
| 102 | 103 | 91 | 101 | 101 | 96 | 100 | 116 | 90 | 98 |
| 114 | 103 | 91 | 101 | 101 | 96 | 100 | 120 | 90 | 98 |
| | | | | | | | | | |
| | | | | | | | | | |
| | 65 | 0 | | 0 | | 105 | 223 | | 0 |
| | 103 | 82 | 101 | 101 | | 0 | 143 | | |
| | | | | | | | | | |
| 130 | 103 | 93 | 101 | 101 | 96 | | 111 | 92 | 103 |
| | 103 | 93 | 101 | 101 | 0 | | 223 | 292 | |
| 100 | 103 | 93 | 101 | 103 | 96 | 100 | 116 | 90 | 98 |
| 102 | 103 | 93 | 101 | 101 | 96 | 100 | 116 | 90 | 98 |
| 102 | 103 | 93 | 101 | 101 | 96 | 100 | 116 | 90 | 98 |
| 140 | 103 | 89 | 101 | 101 | 96 | 100 | 116 | 90 | 98 |
| | 105 | 85 | 101 | 101 | 96 | 100 | 116 | 90 | 98 |
| 102 | 103 | 91 | 101 | 103 | 96 | 100 | 116 | 90 | 98 |
| 102 | 103 | 91 | 101 | 101 | 96 | 100 | 116 | 90 | 98 |
| 118 | 103 | 93 | 137 | 101 | 89 | | 116 | 64 | 98 |
| | | | 81 | | 55 | | | 108 | |
| 102 | 103 | 91 | 101 | 101 | 96 | 100 | 116 | 90 | 98 |
| 102 | 103 | 91 | 101 | 101 | 96 | 100 | 116 | 90 | 97 |
| 102 | 103 | 93 | 101 | 101 | 96 | 100 | 116 | 90 | 98 |
| 102 | 103 | 93 | 101 | 101 | 96 | 100 | 116 | 90 | 98 |
| 102 | 103 | 93 | 101 | 101 | 96 | 100 | 116 | 90 | 98 |
| 102 | 103 | 93 | 101 | 101 | 96 | 100 | 116 | 90 | 98 |
| | | | | | | | | | |
| | | | | | | | | | |
| 102 | 103 | 93 | 101 | 101 | 96 | 100 | 116 | 90 | 98 |
| 138 | 103 | 79 | 86 | 52 | 76 | 90 | 89 | 71 | 98 |
| 157 | 117 | 89 | 96 | 59 | 89 | 100 | 98 | 80 | 112 |
| 157 | 117 | 89 | 96 | 59 | 89 | 100 | 98 | 80 | 112 |
| 157 | 117 | 89 | 96 | 59 | 89 | 100 | 98 | 80 | 112 |
| 157 | 117 | 89 | 96 | 59 | 89 | 100 | 98 | 80 | 112 |

| P24_OtroMat | P25_OtrasMa | P26_Constru | P27_ComercR | P28_TranspC | P29_OtrosSe | P30_OtrosSe | N31_ServNoV | F32_GFCF |
|---|---|---|---|---|---|---|---|---|
| | 104 | 101 | 106 | 89 | 101 | 110 | 124 | 106 |
| | 155 | | 110 | 0 | 135 | 111 | 124 | |
| | | 99 | 103 | 94 | 101 | 103 | 125 | 75 |
| | | | | 94 | 101 | 131 | | 104 |
| | 113 | 91 | 99 | 103 | 101 | 103 | 96 | 67 |
| 105 | 101 | 99 | 106 | 94 | 101 | 110 | 124 | 82 |
| 103 | 104 | 99 | 106 | 94 | 101 | 108 | 124 | 44 |
| 102 | 101 | 99 | 106 | 94 | 101 | 108 | 124 | 48 |
| | | | | | | | | 118 |
| | | | | | | | | 107 |
| | 0 | 198 | 103 | | 117 | | 148 | 587 |
| | 98 | 73 | 103 | | 91 | | 290 | 55 |
| | | | | 58 | | | | 100 |
| 112 | 104 | 101 | 106 | 94 | 101 | 108 | 124 | 44 |
| | 113 | | 106 | 94 | 101 | 108 | 124 | 105 |
| 105 | 101 | 99 | 106 | 94 | 101 | 108 | 124 | 53 |
| 103 | 104 | 99 | 106 | 94 | 101 | 110 | 124 | 78 |
| 103 | 104 | 99 | 106 | 94 | 101 | 110 | 124 | 50 |
| 103 | 101 | 99 | 106 | 94 | 104 | 108 | 124 | 105 |
| 103 | 104 | 99 | 103 | 100 | 101 | 95 | 123 | |
| 103 | 104 | 99 | 106 | 94 | 101 | 108 | 124 | 109 |
| 103 | 104 | 99 | 106 | 94 | 101 | 108 | 124 | 110 |
| 103 | 101 | | 106 | 94 | 98 | 108 | 124 | 109 |
| 103 | 101 | 101 | 106 | 94 | 98 | 110 | 124 | 110 |
| 103 | 104 | 99 | 106 | 94 | 101 | 108 | 124 | 108 |
| 107 | 104 | 99 | 106 | 94 | 101 | 108 | 124 | 109 |
| 103 | 104 | 99 | 106 | 94 | 101 | 108 | 124 | 110 |
| 103 | 104 | 99 | 106 | 94 | 101 | 108 | 124 | 99 |
| 103 | 104 | 99 | 106 | 94 | 101 | 108 | 124 | 110 |
| 103 | 104 | 99 | 106 | 94 | 101 | 108 | 124 | 110 |
| | | | | | | | | |
| | | | | | | | | |
| 103 | 104 | 99 | 106 | 94 | 101 | 108 | 124 | |
| 95 | 86 | 74 | 93 | 64 | 88 | 103 | 99 | |
| 107 | 98 | 84 | 105 | 72 | 98 | 115 | 112 | |
| 107 | 98 | 84 | 105 | 72 | 98 | 115 | 112 | |
| 107 | 98 | 84 | 105 | 72 | 98 | 115 | 112 | |
| 107 | 98 | 84 | 105 | 72 | 98 | 115 | 112 | 0 |

**Figure 5**. Values of the simulated final consumption and GFCF as % of the values of a 42 activity sectors SAM [7] (missing values correspond to 0's in the SAM).